\documentclass[11pt,dvips]{article}
\usepackage{epsfig,times}
\usepackage{floatflt}
\usepackage{picinpar}
\usepackage{moriond}
\usepackage{a4}
\newcommand{\grs}{$\gamma$-rays}
\newcommand{\gr}{$\gamma$-ray\,}
\begin{document}
\title{Supernova Remnants: acceleration of particles and gamma-ray
emission}

\author{H.J. V\"olk}
\address{Max-Planck-Institut f\"ur Kernphysik, Postfach 10\,39\,80, D-69029
Heidelberg, Germany}
\maketitle

\abstracts{
 Particle acceleration in the dynamically evolving environment
of Supernova Remnants is discussed in the framework of a genuinely
time-dependent nonlinear theory, assuming spherical symmetry. As a
consequence the dependence of injection on the angle between shock normal
and external magnetic field direction requires a renormalisation of the
calculated particle fluxes. The recent observational results in TeV
gamma-rays from such objects are discussed and found to be consistent with
theory. We conclude that for the present instrumental sensitivities there
are no reasons to draw premature negative conclusions as to the possible
origin of the Galactic Cosmic Rays below the "knee" in Supernova Remnants.
In addition, theoretical predictions and observations are getting very
close. Therefore the coming generation of ground-based and space-borne
detectors will decide this basic question of astrophysics.
}

\section{Introduction} 

Supernova (SN) explosions in the ensemble of stars constitute the largest,
"steady" mechanical energy input in galaxies. They release an amount
$E_{SN} \approx 10^{51}$~ergs of mechanical energy per event, at a rate of
1 event per 30 to 100 yrs in the Milky Way. Supernova Remnants (SNR) are
also the largest heat source for the Interstellar Gas, and have long been
speculated to be the dominant accelerators of the so-called Galactic
Cosmic Rays (CRs). It is the latter question which we shall discuss here.

 We begin with a review of particle acceleration in the outer SNR shock
wave that communicates the explosion energy to the ambient medium. We
shall in particular discuss the injection process, and will argue that
spherically symmetric models are incomplete without a renormalisation of
the calculated particle fluxes. High energy \gr emission due to inelastic
collisions of the energetic particles with gas nuclei or background
photons is an observational consequence of particle acceleration. We shall
therefore critically evaluate the recently available observational results
in TeV \gr emission from the young objects SN 1006, SNR RX J 1713-3946,
Cas~A, and Tycho's SNR. Several older objects, like IC443 or
$\gamma$-Cygni, are observationally still too complex to draw general
conclusions.

 Finally, we shall turn to the observations of the diffuse \gr emission
from the Galactic disk. At TeV energies it might be interpreted as the
unresolved sum of individual \gr sources in the form of CR sources. These
may be SNRs or other objects, and we may be able to identify them in other
wavelength ranges, like the radio continuum. Thus, apart from a rather
well-known truly diffuse background contribution, the spatial distribution
of this "diffuse" TeV \gr emission should be compared with the known SNR
distribution in the Galaxy and, mutatis mutandis, its energy spectrum
ought to correspond to the inferred \textit{average} CR source spectrum.
\section{Particle acceleration in SNRs} 

 The \grs\, from SNRs stem from particles that have at some time in the past
been accelerated at the outer SNR shock to a momentum distribution that is
nonthermal; it corresponds to a power law $\propto p^{- \beta}$. In
fact, the number density $n_i$ of particles participating in the
acceleration process, the injection fraction, is quite small compared to
the thermal particle density $n_{th}$, with $\eta = n_i/n_{th} \approx
10^{-4}$. In one plausible picture \cite{bennett}$^,$ \cite{malkovV95}$^,$
\cite{malkov98}$^,$ \cite{malkovV98} - see however \cite{scholer} for a
different view - the injected ions are those which are able to escape
from the shocked (i.e. suddenly decelerated and heated) downstream thermal
distribution along the magnetic field into the incoming plasma region
upstream of the shock, where they excite MHD waves (``Alfv\'{e}n waves'') and
thus initiate the diffusive acceleration process.


 These scattering waves \cite{bell} keep the particle distribution close
to isotropic, and in the process particles are stochastically scattered
back and forth across the shock front many times, gaining energy in each
shock crossing, before being convected downstream together with the
scattering wave field. This picture holds in its simplest form for
nonrelativistic shocks, i.e. for the case in which the difference in flow
speed across the shock front is nonrelativistic. In any conventional view
SNR remnants fulfill this condition, and in the sequel we shall confine
our attention to this case.


\begin{figwindow}[1,l,{\includegraphics[width=5cm]{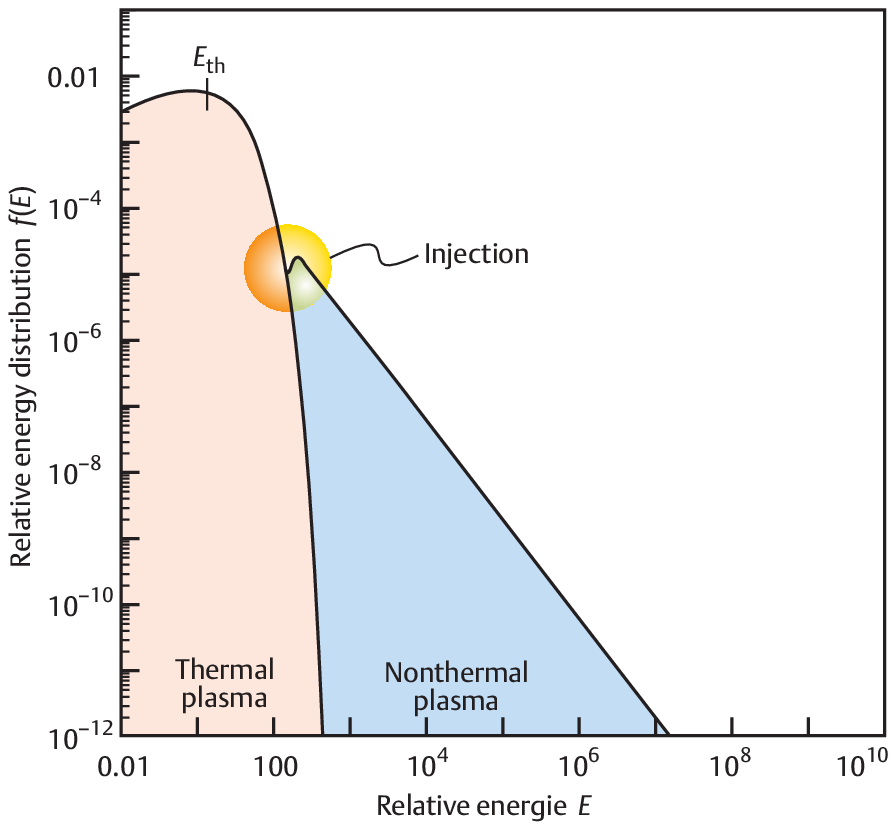}},
{ Particle energy distribution at the shock
(arbitrary units). The downstream thermal plasma has
a quasi-Maxwellian distribution with thermal energy
$E_{th}$, whereas above a somewhat larger energy
(injection) the accelerated nonthermal distribution
starts to dominate. \label{fig:injection}}]
 Apart from the fundamental aspect of the excitation of scattering waves
by the accelerating particle component itself, shock acceleration can also
be described in terms of a well-known transport equation for the isotropic
part of the distribution function that contains the mechanisms of
adiabatic energisation/deceleration, spatial diffusion and convection.  
This diffusive shock acceleration process has been reviewed extensively in
the past, e.g.  
\cite{drury}$^,$ 
\cite{blandford}$^,$ \cite{berezhkoK}. 

 In order to go beyond the test particle limit, it is apparently
necessary to include the energy and momentum exchange of the scattering
particles with the thermal gas (plasma) and the wave field in the
calculation of the \textit{overall} dynamical evolution of the system.
Then, for a strong shock, the downstream nonthermal and thermal energy
densities turn out to be comparable, $E_{nonthermal} \sim E_{thermal}$.
This means that the process can be highly nonlinear and 
\mbox{efficient \cite{malkov97};} for a very recent review, see \cite{malkovD}. The
differential energy distribution in the relativistic range is close to a
$E^{- 2}$~-law, i.e. it contains essentially equal energy per decade in
particle energy $E$. The ratio of the energy densities and the spectrum
are both rather close to what is observed for the CRs in our Galaxy and in
many astronomical objects which exhibit nonthermal emission. 
\end{figwindow}
\newpage
\begin{figwindow}[1,l,{
\includegraphics[width=5cm]{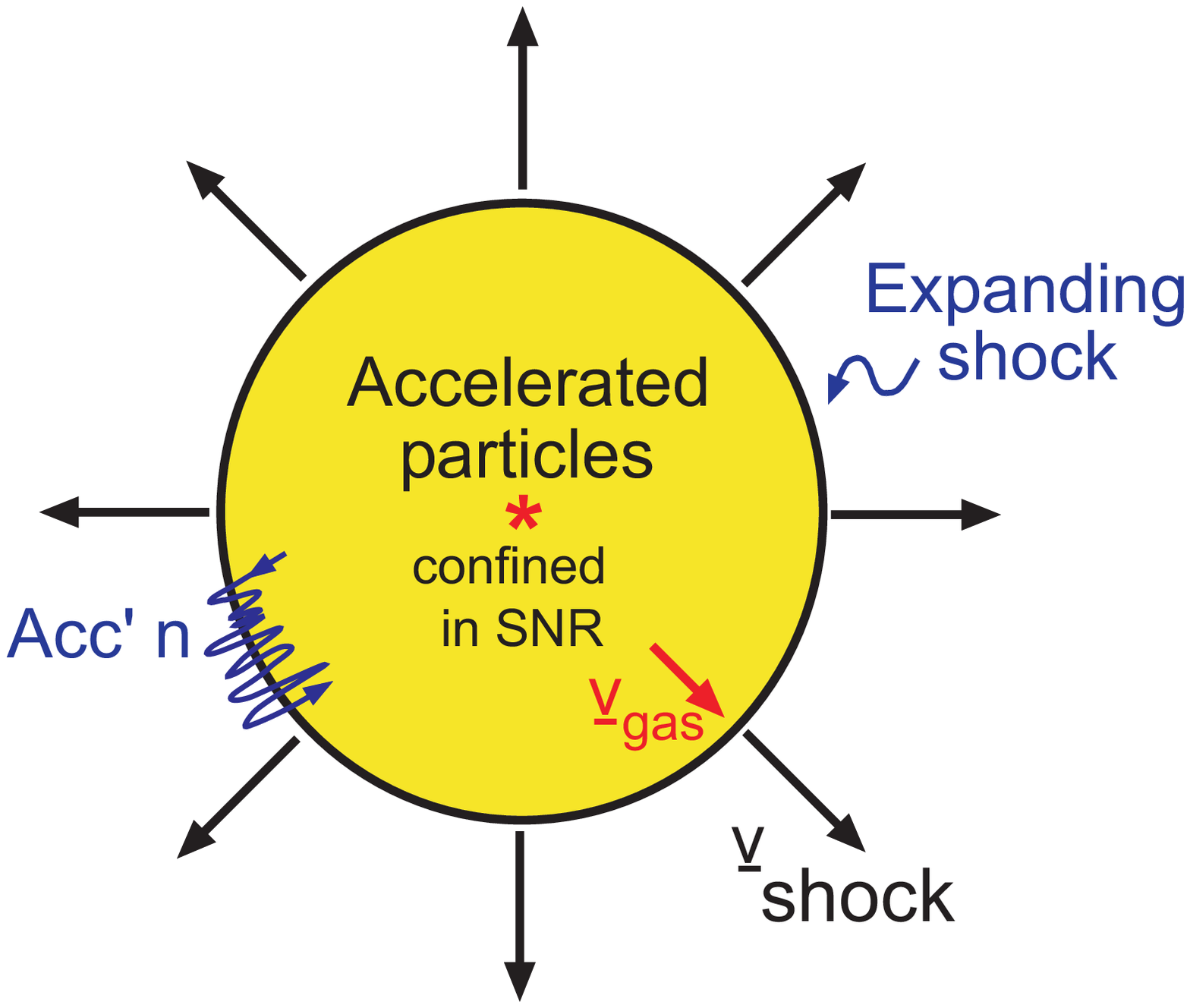}},
{ At the spherical SNR particles are injected from the inside and remain
diffusively confined after acceleration, expanding with the thermal gas
until its velocity $v_{gas}$ decreases below characteristic ambient
velocities.\label{fig:acceleration}}]
All this pertains to nuclei which are known to dominate electrons in the
Galactic
 CRs by a factor of $\sim 100$. However, the large Thompson cross section
allows the electrons to radiate very effectively, and therefore they play
a significant role for the nonthermal emission from SNRs
\cite{mastichiadis} and other nonthermal sources.

 The special characteristics of SNRs lie in the fact that they are the
result of a strong point explosion. Therefore they are intrinsically time
dependent objects, to lowest order spherically symmetric, and indeed
strongly nonlinear accelerators. Particles that have been accelerated
remain inside the expanding remnant, undergoing adiabatic expansion losses
and diffusive transport in the interior, before being ultimately released
into the Interstellar Medium  when the remnant gets old and starts to
decay (Fig. \ref{fig:acceleration}).
\end{figwindow}
\subsection{Volume-integrated particle, \gr spectra, and energies} 
\begin{figwindow}[2,r,{
\includegraphics[width=5cm]{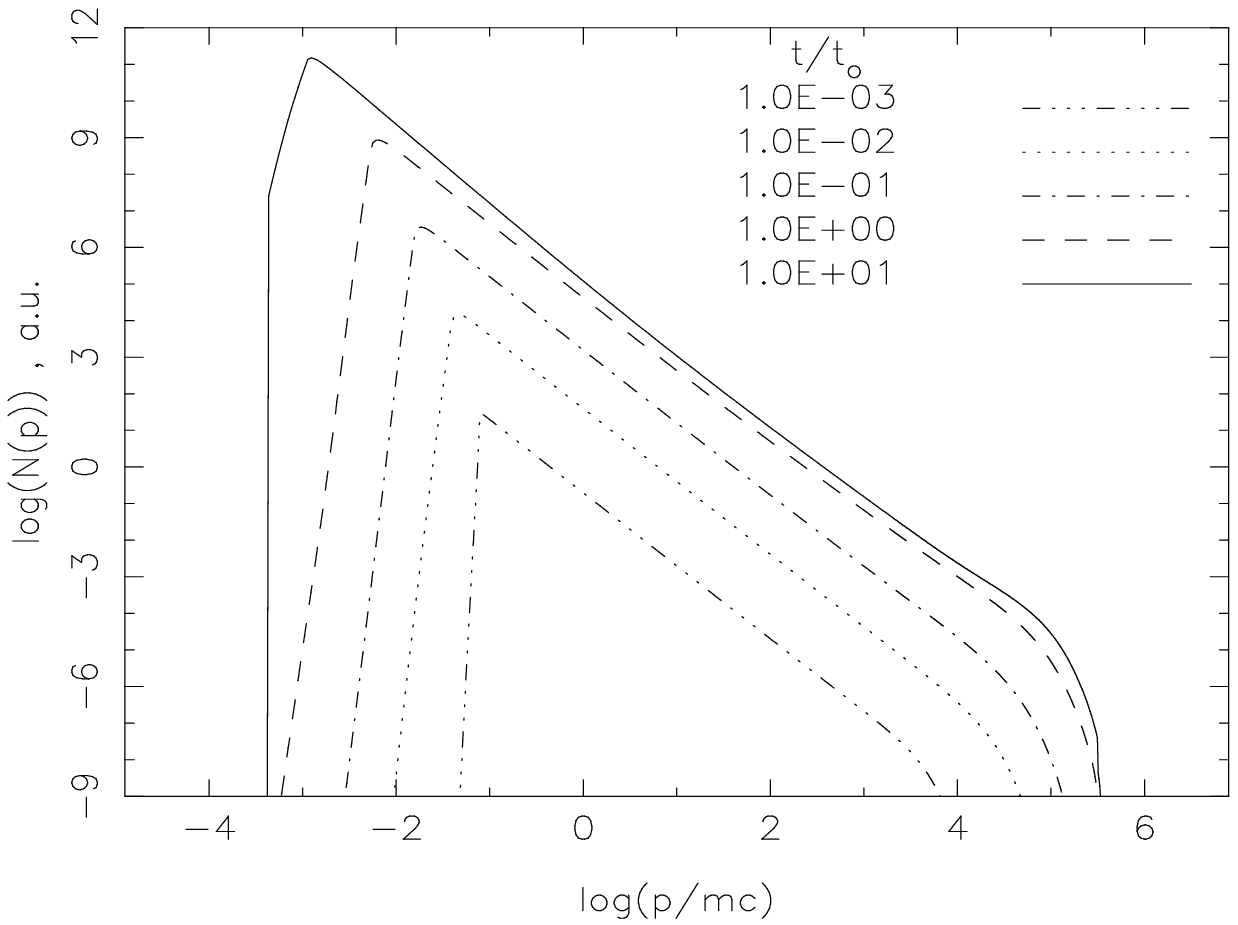}},
{(a) Volume-integrated proton momentum spectra $N(p)$ for
different SNR ages $t$ in units of the sweep-up
time $t_0 = 1893$~yr, for an injection fraction $\eta =
10^{-4}$, upstream density $n = 0.3$~H-atoms
cm$^{-3}$, $E_{SN} = 10^{51}$~erg, ejected mass
$M_{ej} = 10 M_{\odot}$, and $B = 5 \mu$G.
\label{fig:protons}}]
 In the simplest case of a uniform Interstellar Medium, the external
magnetic field $B$ is uniform and thus the angle $\Theta_{nB}$ between the
shock normal and $B$ varies systematically over the shock surface.
Assuming for simplicity nevertheless spherical symmetry for the solution
of the nonlinear acceleration problem, not only the acceleration process
is calculated for a "parallel shock" ($\Theta_{nB}=0$), but also the
injection efficiency is assumed to be the same over the entire shock
surface, indeed equal to that for $\Theta_{nB}=0$. Whereas the first
approximation is possible for almost all values $\Theta_{nB}\neq 0$
, e.g. \cite{drury}, this is not true for the injection rate, e.g.
\cite{baring}$^,$ \cite{malkovV95}, and we shall argue that this requires a
renormalisation of the spherically symmetric result. 
 For the case of a uniform ambient medium, typically appropriate for SNe
type Ia and core collapse SNe type II from precursor stars considerably
less massive than $20 M_{\odot}$, the nonlinear set of equations
describing nucleon acceleration and the dynamical backreaction on the
thermal gas has been solved numerically
\cite{berezhko94}$^,$ \cite{berezhkoV97}, covering
the complete time evolution. Fig. \ref{fig:protons} shows an example of
the momentum distribution of energetic protons, spatially integrated over
the SNR volume \cite{berezhkoV97}.
\end{figwindow}
\begin{figwindow}[3,l,{
\includegraphics[width=5cm]{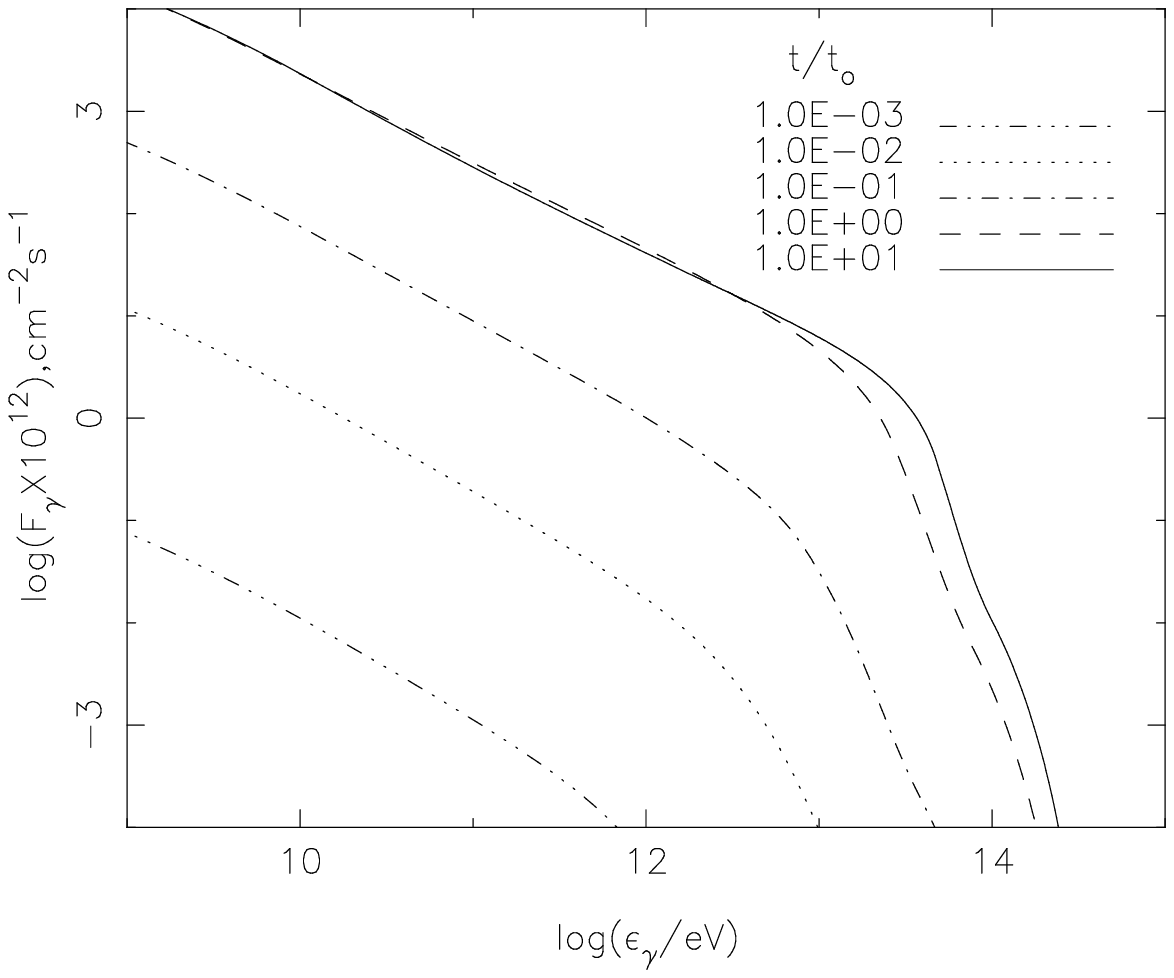}},
{Volume-integrated $\pi^0$-decay \gr spectrum corresponding to the parameters of Fig. \ref{fig:acceleration}.
\label{fig:gamma_rays}}] 
 Above the injection momentum the integrated spectrum is close to a power  
law which is getting harder towards the cutoff due to nonlinear shock
modification; the overall average spectral index is close to 2; the
maximum momentum reaches about $3 \times 10^{14} {\rm eV}/c$ at late times
for the particular parameters chosen \cite{berezhkoV97}. (As a side remark: radio synchrotron
electrons, with typical energies as low as 10 GeV, should therefore have a
somewhat steeper spectrum for the nonlinearly modified, very strong shocks
characterising young remnants \cite{ellisonR}$^,$ \cite{duffy}; in fact, the 
observed radio spectral index often exceeds the value 0.5 that
would correspond to the spectra at high momenta)

 Of similar interest is the volume-integrated integral \gr spectrum from
the SNR due to $\pi^0$-decay from inelastic collisions with thermal gas
nuclei, Fig. \ref{fig:gamma_rays}.

For a spatially resolved observation one has, of course, to compare the
locally calculated emission.
\end{figwindow}
\begin{figwindow}[1,l,{
\includegraphics[width=5cm]{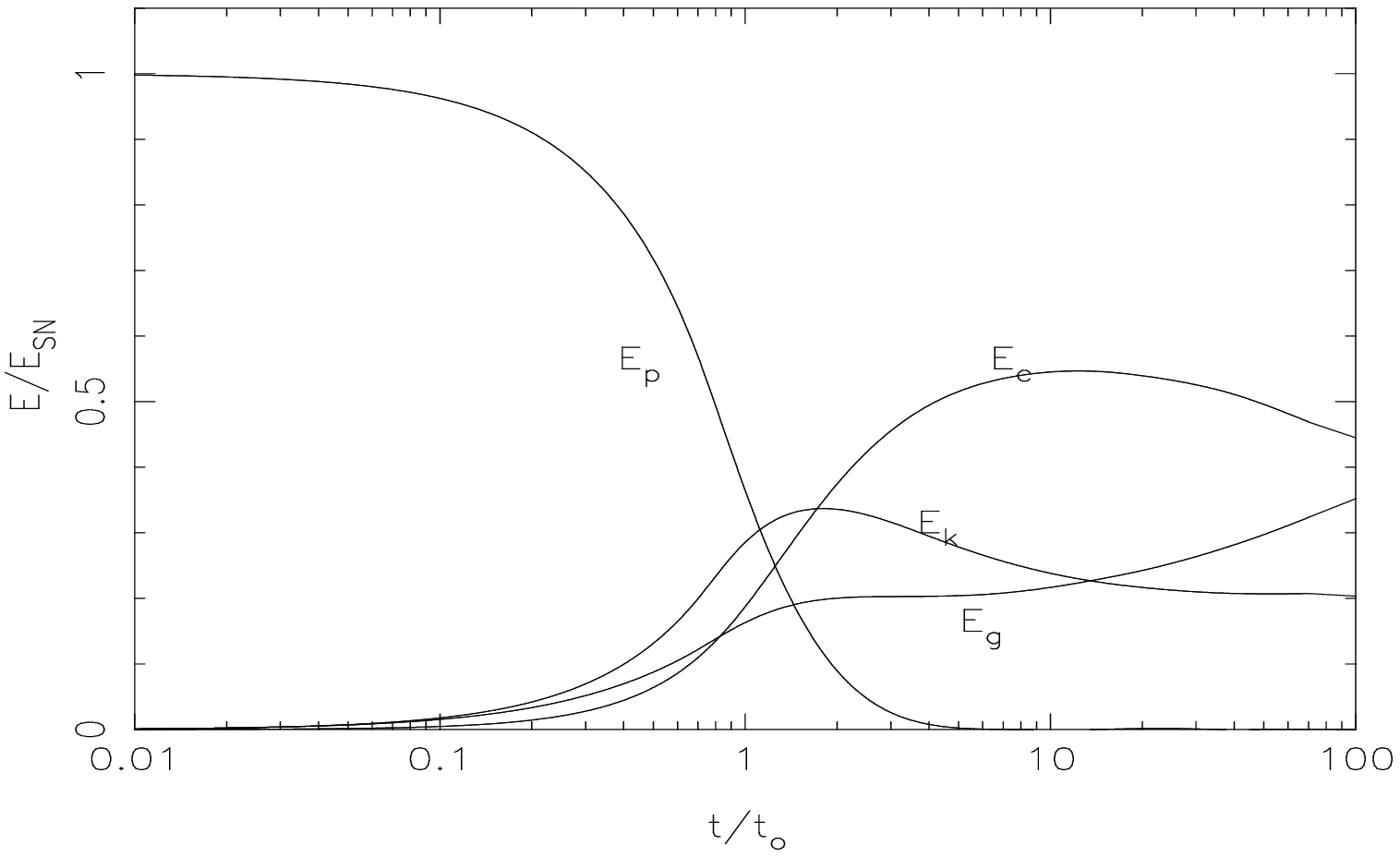}},
{Volume-integrated energy fractions for the parameters of Fig.
\ref{fig:protons}. The quantities $E_p$, $E_g$, $E_c$, and $E_k$ correspond to ejecta
energy, gas thermal energy, CR energy, and gas kinetic energy,
respectively.
\label{fig:energies}}]
Another aspect of the time-dependent evolution can be recognised in the
different components of the volume-integrated energy in the system,
Fig. \ref{fig:energies}. Whereas initially the entire hydrodynamic
energy is in the ejected mass, the thermal and kinetic energies of the gas
and the energy $E_c$ in accelerated particles increase as a function of
time. $E_c$ is greater or about equal to the other energy components in
the Sedov phase, reaching a value $E_c/E_{SN} \approx 0.5$ in the late
Sedov phase, when particles should be released \cite{berezhkoV97}.

 CR observations on the other hand suggest a significantly lower
\textit{average} source efficiency. Indeed, according to standard estimates
 \cite{berezhkoBG}, the energy input into CRs from an individual SNR
is about 
$(dE/dt)_{Sources} \approx 0.1 \times \nu_{SN} E_{SN}$, 
and we shall use this number in the sequel.
\end{figwindow}

\parbox{\linewidth}{\rule{0pt}{.8cm}}
\subsection{Renormalisation of integral \gr fluxes} 
\parbox{\linewidth}{\rule{0pt}{.8cm}}

 The physical reason for the overproduction of nonthermal energy in the
spherically symmetric model lies in its assumption of a constant injection
efficiency over the shock surface. This is actually not the case. To
lowest order the mean magnetic field configuration for a point-like energy
input into an environment with a uniform field looks like the cartoon in
Fig. \ref{fig:renormalisation}: The external magnetic field lines are
refracted into the interior. The minimum field-aligned speed of a
downstream particle required to outrun the radially propagating shock wave
into the upstream plasma along a field line is proportional to
cos$^{-1}\Theta_{nB,2}$ and appears in the (quasi-exponential) tail of the
velocity distribution of the shocked downstream plasma.
\newpage

\begin{figwindow}[1,l,{ \includegraphics[width=5cm]{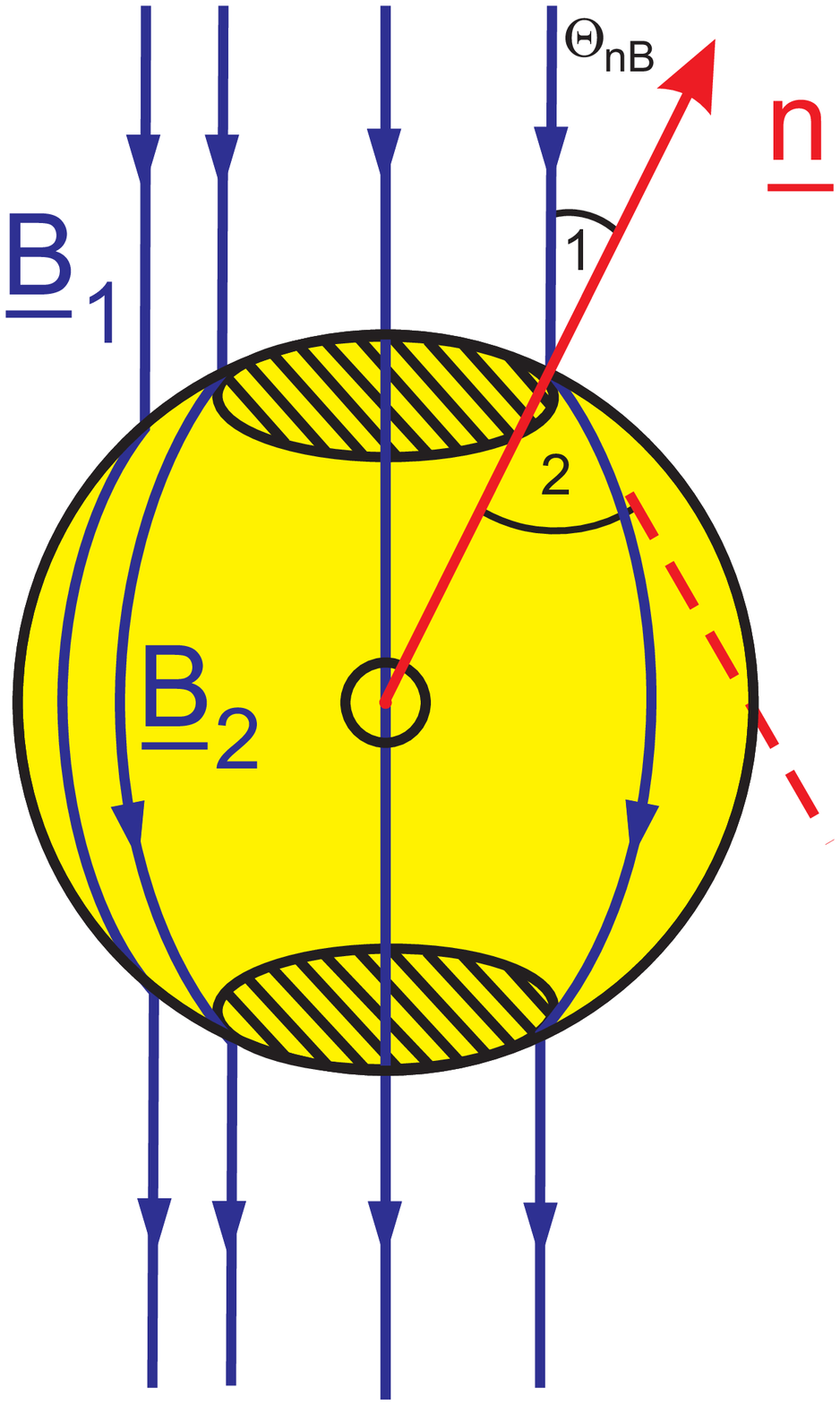}}, 
{Cartoon of the B-field refraction in a SNR situated in a uniform field
region. The angle $\Theta_{nB,2}$ between the downstream field $\underline{B}_2$ and
the shock normal
direction $\underline{n}$ becomes so large beyond the hatched "polar"
region that injection is effectively inhibited.
\label{fig:renormalisation}}] 
 In reality, the magnetic field in the collisionless (sub)shock is much
more complicated, even in the "parallel" case $\Theta_{nB,1} = 0$
\cite{bennett}$^,$ \cite {malkov98}. As a rough approximation we
may nevertheless scale the injection velocity with the factor 
cos$^{-1}\Theta_{nB,2}$ which dramatically reduces the injection rate for
increasing shock obliquity.

 We shall not go into the details here. However, it is clear that for a
sizeable fraction of the shock surface near the "equator",
cos$\Theta_{nB,1} = 0$, particle acceleration will be inefficient relative to
that in the regions near the "poles". 

 To lowest order, the spatial integral of the accelerated spectrum should
therefore be reduced by multiplying the spherically symmetric result with
a \textit{renomalizing factor} given by the ratio of 
shock 
surface area with
efficient injection to the total shock area. Instead of such a theoretical
factor we shall use here an even simpler prescription. This is an
\textit{empirical} renormalization factor, on the assumption that SNRs are
indeed the sources of the Galactic CRs, and corresponds to the ratio of
the expected value $E_{c,\, observed}/E_{SN} \approx 0.1$ to the calculated
value of $E_c/E_{SN}$. For the case of Fig. \ref{fig:energies} it is
about 1/5.
\end{figwindow}

\section{Observed young SNRs}
\subsection{SN 1006 in the Southern Hemisphere}

 Being a bright radio synchrotron source, this historical remnant of a SN
type Ia has also been detected in nonthermal X-rays \cite{koyama95}$^,$
\cite{allen97}, believed to be synchrotron emission. The X-ray morphology
is characterized by two symmetrically situated emission regions of unequal
strength, reminiscent of the "polar" regions of preferred injection
discussed before. Given the implied high electron energies of $\sim 100$~
TeV \cite{reynolds} and assuming magnetic fields below $10~ \mu$G , SN
1006 was consequently also predicted as an Inverse Compton (IC) \gr source
by \cite{pohl}$^,$ \cite {mastichiadisJ}$^,$ \cite{yoshida}. A TeV \gr
detection was finally reported by the Japanese/Australian CANGAROO
collaboration \cite{tanimori}. The report contained also indications of a
\gr morphology resembling that in nonthermal X-rays.

 Using the popular estimate \cite{dav}$^,$ \cite{naito} of the expected $\pi^0$-decay
emission for the parameters of SN 1006, it turns out \cite{voelk97} that
the CANGAROO flux is more than a factor of about four higher than this
estimate, making a hadronic \gr source implausible for the given
parameters. A detailed recent kinetic modeling of the nonthermal X-ray and
\gr emission \cite{berezhkoKP} raises the $\pi^0$-decay luminosity by a
factor $\sim 2$, making it comparable to the IC \gr luminosity. However,
as argued above, this flux should be renormalized - by a factor of $\sim 6$ -
which once again appears to rule out a hadronic origin. \cite{aharonianA}
have presented a more phenomenological discussion which in particular
points out that the implied \gr morphology fits much better a shell-type
hadronic \gr emission rather than an IC emission that they expected to be
rather uniform across the remnant, given the uniformity of the target
photon distribution. On the arguments from Fig.
\ref{fig:renormalisation} however, a nonuniformity of the \gr emission
is expected in any case since the particle distribution is asymetric.

 Given the present experimental results, much depends on the strength of
the effective magnetic field in the acceleration region. There is clearly
a need to observe at lower and higher \gr energies than 1 TeV with more
sensitive instruments to obtain precise spectral information, especially
about \gr cutoffs, in order to allow a distinction between a nucleonic and
an IC origin of the $\gamma$ radiation. At the same time, the available
improvements in spatial resolution will be important in order to obtain a clear
\gr morphology for this large remnant of $\approx 0.5^{\circ}$ diameter.

\subsection{SNR RX J1713.7-3946 in the Southern Hemisphere} 

 This large, $1^{\circ}$ diameter SNR had been discovered in the
\textit{ROSAT} All-Sky \mbox{Survey \cite{pfeff}} and was observed as a strong
nonthermal X-ray and radio source \cite{koyama97}$^,$ \cite{slane}. In fact,
SNR RX J1713.7-3946 is the only SNR that does not show significant
evidence for thermal X-ray emission from any portion of the remnant! It has been
reported as a TeV \gr source by CANGAROO \cite{muraishi}. However, the
parameters, like distance, circumstellar environment, and age, are not
well known. In particular, the distance has been estimated as widely
different as 1 kpc \cite{koyama97} and 6 kpc \cite{slane}, at gas
densities of $0.28$~H-atoms cm$^{-3}$. The latter authors also concluded
that the progenitor should have been a massive star, in whose wind-blown
bubble the SNR shock still propagates today.

 If the reported \gr emission is interpreted as Inverse Compton emission
in the Cosmic Microwave Background, then it is consistent with a rather
low magnetic field strength of $\approx 11 \mu$G, \cite{muraishi}
independently of distance, etc., given the observed nonthermal X-ray flux
as synchrotron emission from the same electrons. Whether the \gr emission
could also be due to $\pi^0$-decay is an open question, given all the
uncertainties. However, if the remnant was indeed as distant as 6 kpc,
then the $\pi^0$-decay flux would be negligible compared to the observed
flux. If, in addition the SNR shock still propagated in the rarefied wind
bubble of a massive star, then the $\pi^0$-decay flux should be
significantly reduced due to the dilution effect in the bubble
\cite{berezhkoV00}. A lot more work must be done before this extraordinary
source is clearly understood.

\subsection{Tycho's SNR in the Northern Sky}

 The progenitor of this SN type Ia was probably situated in a more or less
uniform environment, making it another ``astronomically simple'' object,
like SN 1006 in the South. Although the observed X-ray continuum between
10 and 20 keV \cite{petre} from RXTE may be interpreted as synchrotron
emission (see however \cite{laming98} for a different view), the
X-ray flux is in general dominated by line emission which suggests that IC
\gr emission is not the dominant contributor to the expected TeV \gr flux.
In many ways Tycho should be the prototype of a recently born CR source in
the Galaxy. An earlier attempt to detect it was made by the Whipple group
which, after an observation time of $\sim 14$ hrs, could set an upper
limit to the flux at 300 GeV \cite {buckley}. More recently the HEGRA
stereoscopic system has observed the source for $\sim 65$ hours at energies above
\mbox{1 TeV \cite{aharonianR}}. No significant \gr flux was detected either, leading
to an upper limit of $5.78\times 10^{-13}$ph cm$^{-2}$ sec$^{-1}$ above 1
TeV, roughly 4 times lower than the Whipple upper limit if the different
energy is accounted for. If the radio and the keV flux are interpreted as
synchrotron radiation, then the non-observation of a corresponding IC \gr
flux implies a lower limit to the magnetic field strength of about $20
\mu$G. Using the analysis of the ASCA detection \cite{hwang}, on the
other hand, weakens such an upper limit to about $6\mu$G. Although on the
high side for an unperturbed upstream interstellar field, even a $20\mu$G
field strength might actually exist, a $6\mu$G field is in any case
possible .

\begin{figwindow}[1,l,{\includegraphics[width=7.02cm]{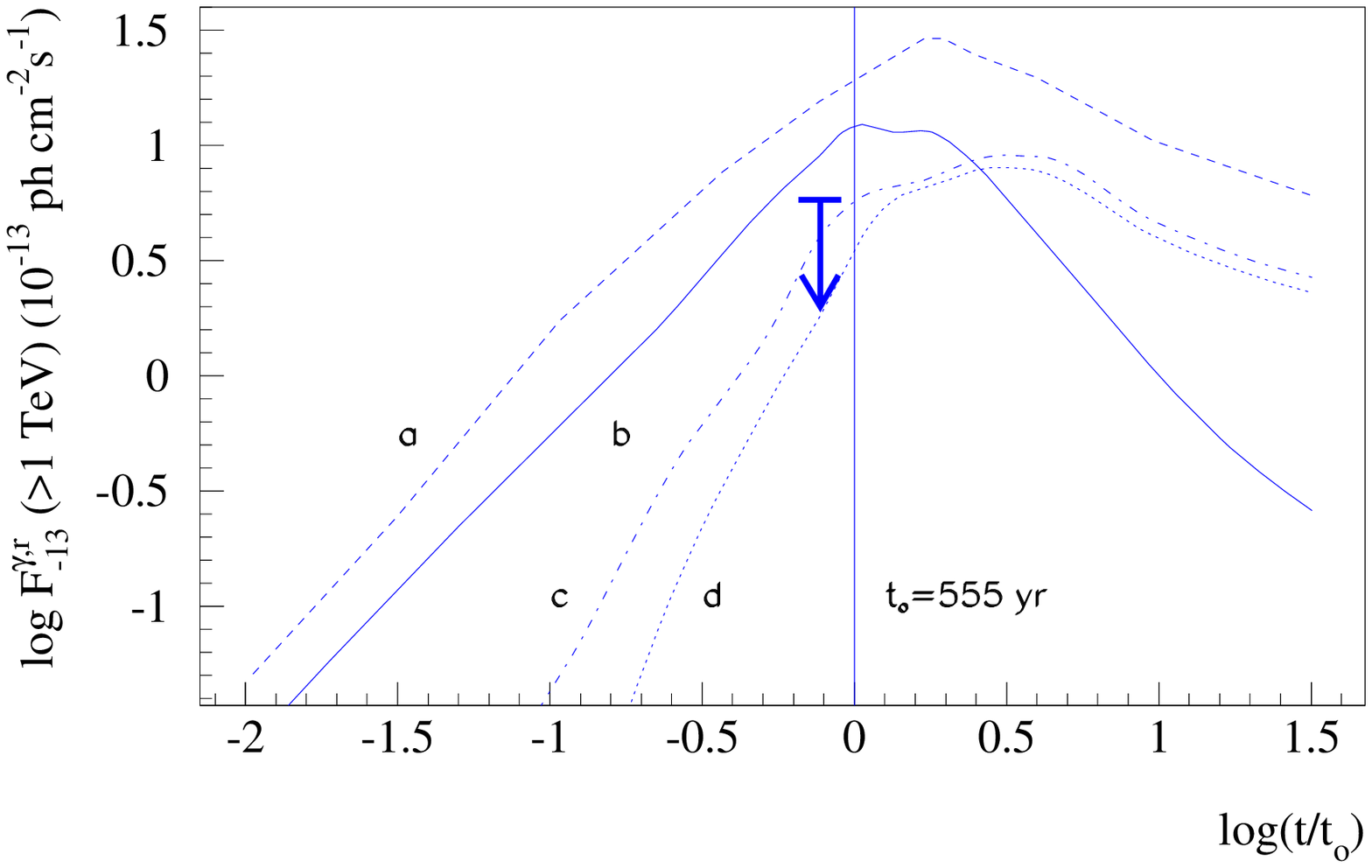}},
{Time dependent calculation of the $\pi^0$-decay \gr flux for Tycho's SNR 
for four different parameter sets: dashed line, case a) $\eta =
10^{-3}$ and $B = 5 \mu$G; solid line, case b) $\eta = 10^{-3}$ and
$B= 30 \mu$G; dash-dotted line, case c) $\eta = 10^{-4}$ and $B = 5
\mu$G; dotted line, case d) is a single velocity ejecta case with
$\eta = 10^{-4}$ and $B = 5$
\label{fig:Tycho}}]
 The hadronic \gr flux predictions \cite{aharonianR}, rescaled from \cite{berezhkoV97}
 and renormalised cf. section 2, give a result that is
equally close to the deduced upper limit:
 The experimental results appear however to exclude high injection rates,
$\eta = 10^{-3}$, cases (a) and (b) in Fig. \ref{fig:Tycho}.
Disregarding the rather unphysical case (d) that assumes a constant mean
ejecta velocity, the upper limit is only slightly
above the low injection case (c), with $\eta = 10^{-4}$.

 Therefore, the \gr observations of Tycho's SNR come so close to
theoretical predictions that a deeper observation, like it will be
possible with the VERITAS array or the MAGIC telescope, should indeed lead
to a detection, even taking into account the general astronomical
uncertainties in distance, ambient density and total energy that exist also for this
object.
\end{figwindow}

\subsection{Cassiopeia A in the Northern Hemisphere}

\begin{figwindow}[2,l,{
\includegraphics[width=6.99cm]{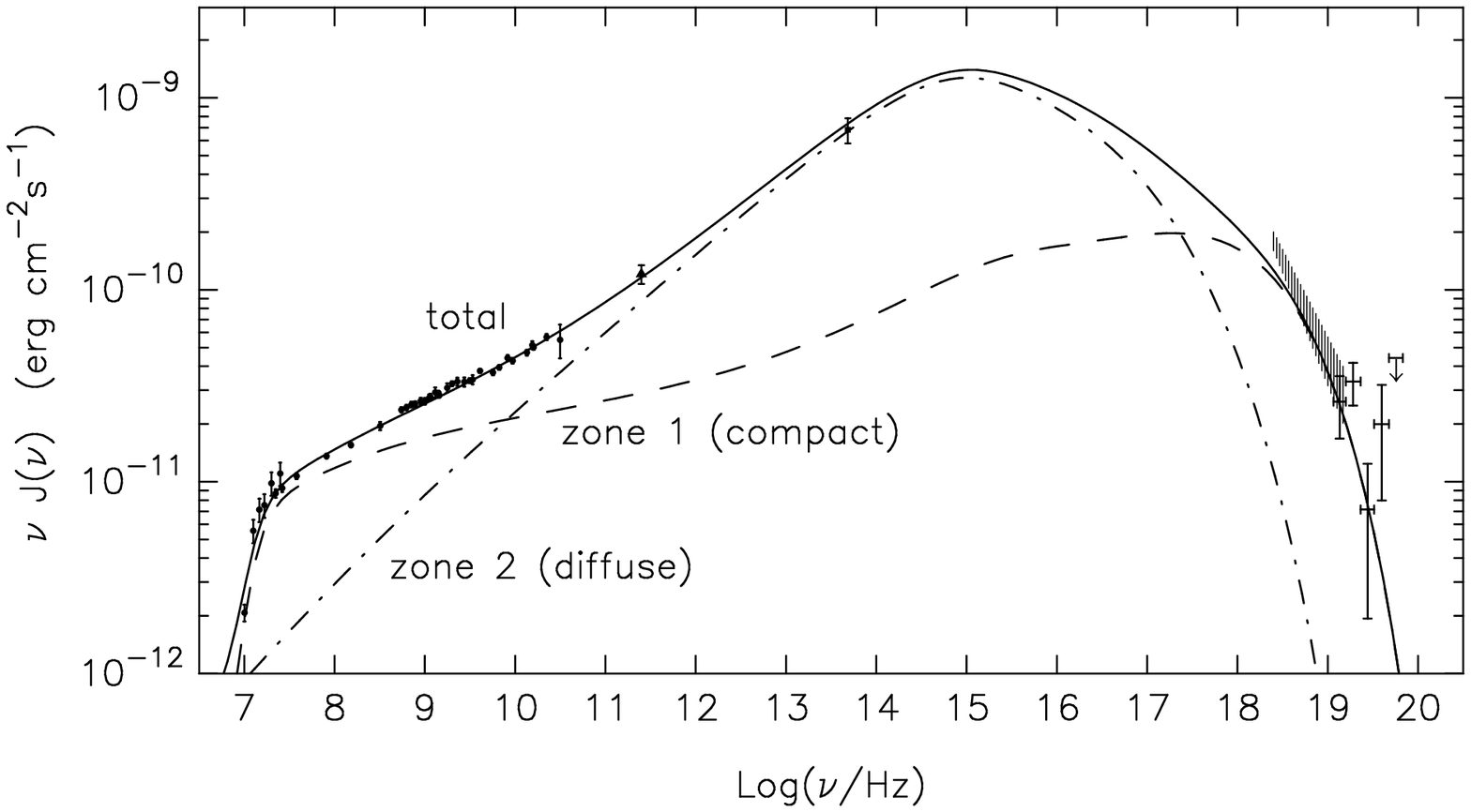}},
{Model for the synchrotron SED for Cas A. The inhomogeneous remnant is
divided into 2 components (zones), where zone 1 comprises the many
small-scale features, assumed to be sources of energetic particles, and
the remaining diffuse emission region (zone 2) fed by zone 1 sources and
the outer SNR shock. 
 A good fit to the data points is
obtained
\label{fig:casa_synchrotron}}]
 Cas A is presumably the youngest known SNR in the Galaxy, dating back to
about 1680, and the strongest radio source in the sky. It has been    
detected in TeV \grs\, in the deepest \gr observation up to now
\cite{aharonianP} at a level of $\sim 3$~\% of the Crab flux. The   
observations are described in these Proceedings \cite{horns}. If besides
the radio continuum also the hard X-ray flux \cite{allen97} is interpreted
as synchrotron emission \cite{atoyanATV}, see \mbox{Fig.
\ref{fig:casa_synchrotron}},
a corresponding \gr emission due to Bremsstrahlung and IC scattering
should result. The existing observational results are compared with model
predictions in \mbox{Fig. \ref{fig:casa_gr}}. The magnetic field strengths
assumed are in the \mbox{$\sim 1$ mG} range, whereas the implied matter density
corresponds to 15 H-atoms cm$^{-3}$. A total energy in protons $W_p = 2
\times 10^{49}$~erg was assumed, for a proton differential spectral index
of 2.15 and a cutoff energy of \mbox{100 TeV}. The higher extension in energy,
expected for the nucleonic spectrum, slightly favors a nucleonic origin
for the \gr emission.
\end{figwindow}
\newpage
\begin{figwindow}[2,l,{
\includegraphics[width=7.35cm]{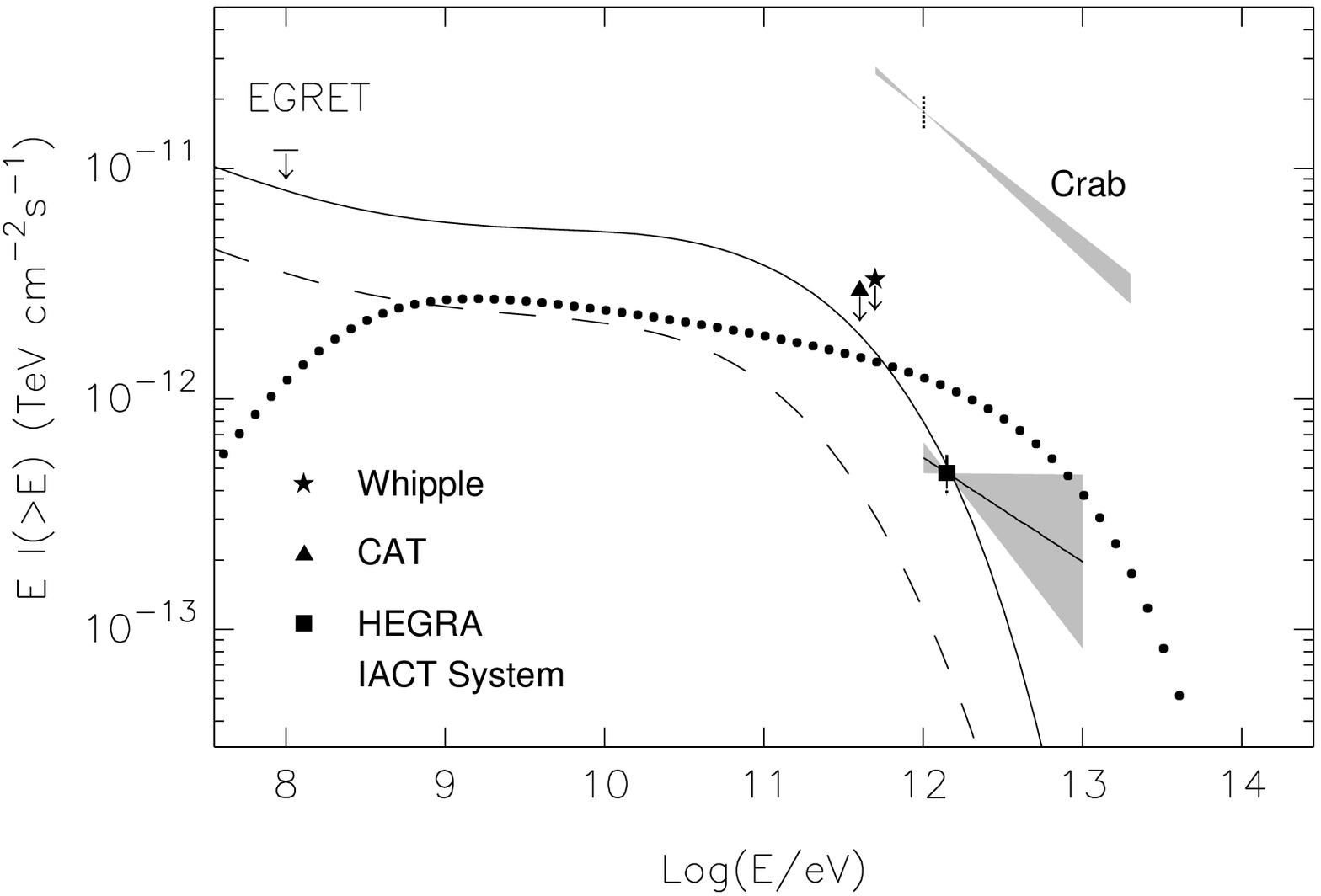}},
{Measured flux and spectral index of Cas A in the context of model
predictions. Shaded area shows 1$\sigma$~error range for the spectral
index. Solid and dashed lines show predicted IC plus bremsstrahlung flux
for two different values of the B-field, whereas the dotted line shows a
possible form for the ha\-dron\-ic \gr flux. Also indicated are the upper
limits measured by EGRET, Whipple and CAT, and the values for the Crab
Nebula.
\label{fig:casa_gr}}]
 Recently, Laming \cite{laming01} has argued that the hard X-ray emission
could also be Bremsstrahlung from nonthermal electrons with several tens
of keV, energised by lower hybrid waves from shock reflected ions at
qua\-si-per\-pen\-dic\-ular shocks. Such a mechanism had been first suggested by
\cite{galeev}. Since the progenitor of Cas A is presumably a massive,
fast-ro\-ta\-ting Wolf-Rayet star in whose Red Supergiant wind the SNR shock
is presently propagating \cite{borkowsky}, the shock may indeed be largely
qua\-si-per\-pen\-dic\-ular. If this latter electron energisation scenario
produces enough energy to yield the observed Bremsstrahlung, the detection
of Cas A in \grs\, would necessarily imply a dominant hadronic $\gamma$-ray
pro\-duc\-tion! Moreover, the effect opens the Pandora box also for SNe of the
type Ia like Tycho, or even SN 1006, in that a relevant fraction of the
observed hard X-ray continuum could be nonthermal bremsstrahlung due to
the partly quasi-perpendicular nature of the SNR shock in a uniform
external magnetic field (section 2.2). This would clearly shift the
interpretation towards an increasing nucleonic \gr fraction in general.
\end{figwindow}

\section{Older core collapse SNRs in TeV $\gamma$-rays}
Several older core collapse SNRs have been observed as well, and none
of them could be detected in TeV \grs\, \cite{buckley}$^,$ \cite{hess}$^,$
\cite{rowell}, even though they had been detected in GeV \grs\, by the EGRET
instrument. We shall concentrate here on $\gamma$-Cygni and IC
443, since both the Wipple and the HEGRA telescopes have observed them.
Even if the upper limits would have corresponded to detections, a naive
straight line interpolation between the EGRET and the TeV fluxes would not
correspond to the expected CR source spectrum but rather to one that is
considerably steeper.

 In the case of $\gamma$-Cygni also the EGRET error circle  is fully inside and much smaller than the radio shell that marks
the SNR morphology. Therefore the EGRET source cannot be the shell SNR. In
fact, later X-ray measurements \cite{brazier} have indicated the presence
of a young pulsar, although the discussion is not really closed
\cite{voelk97}.

 IC 443 is a more complex case. In contrast to $\gamma$-Cygni, this remnant
appears indeed to be interacting with dense cloud material, as indicated
by the presence of OH maser emission \cite{claussen}. However, the X-ray
emission is again characterised by localised patches, partly outside the
EGRET error circle, and not delineating a uniformly illuminating
if inhomogeneous shell remnant. The expected $\pi^0$-decay \gr flux is,
within the considerable uncertainties of the astronomical parameters,
about equal to the Whipple and HEGRA upper limits. Therefore a deeper \gr
observation might well lead to a TeV detection.

 From the point of view of acceleration theory both SNRs might possibly
also be rather rare and older so-called Wind-Supernovae, where the SNR
shock is outside the stellar wind region of the massive progenitor but
still inside the low-density, hot bubble of shocked wind material. It is
expected that such objects exhibit a low \gr emission of nucleonic origin,
and presumably a low IC/Bremsstrahlung emission as well
\cite{berezhkoV00}.

 Even though the experimental situation is not fully clarified, these sources
are complex enough in every respect that it is difficult to draw any firm
conclusions at present. This has not prevented rather pessimistic reactions,
either observationally \cite{buckley}, or theoretically \cite{kirk}. The
prime lesson we can learn from these examples is probably that SNRs have very
individual characteristics, far from being simple templates like low-mass
main sequence stars.

\section{Diffuse \gr emission from the Galactic disk}

 EGRET observations have shown an enhanced \gr emission from the Galactic
disk above a few GeV in comparison with the standard models which
otherwise describe the spectral intensity and \gr morphology of the disk
fairly well \cite{hunter}. This hardening of the energy spectrum, which is
particularly pronounced near the disk's midplane, might be
due to CR propagation effects, e.g. \cite{aharonianAt}$^,$ \cite{voelk00}.
An interesting alternative is its origin in the hard spectrum of
\gr production \textit{inside} SNRs if these SNRs contribute a
sufficiently strong unresolved background of CR sources in the disk.
Estimates \cite{berezhkoBG} have shown that this should indeed be the
case, the source contribution dominating above $\sim 100$~GeV.

 Recent HEGRA observations have not detected a diffuse \gr background in
the Galactic disk above 1 TeV \cite{aharonianL}. However, they have been
able to establish an upper limit which lies only a factor of $\sim 2$
above the predicted emission from the unresolved ensemble of CR sources in
the form of SNRs. If this was indeed the case, then deeper observations
with the upcoming arrays H.E.S.S. and CANGAROO in the Southern Hemisphere
should be able to detect this background and thus to establish the
\textit{average} form of the Galactic CR source spectrum due to SNRs.
Furthermore, a TeV-detection of the radial \gr distribution in the Galaxy
would be a complementary result, allowing a direct comparison with the SNR
distribution inferred from radio measurements.

\section{Conclusions}

 SNRs are complex nonthermal objects, often situated in a disturbed
environment, and they defy oversimplified quantitative theoretical as well
as -- in many cases -- observational interpretations. On the other hand,
they are the only nonthermal astrophysical sources for which something
exists what one might call a real theory in the first place, even if it is
not complete as we have seen.  The sensitivity of present \gr detectors is
still marginal for the detection of such objects \cite{dav}. We have
argued that the existing, in their majority unsuccessful, observational
searches are nevertheless consistent with theoretical expectations. For
the detected remnants it is difficult to seperate the \gr fluxes of
nucleonic and of electronic origin. For this reason, and for the fact that
none of the three claimed detections has yet been confirmed by independent
measurements, the question of a SNR origin of the Galactic CRs below the
"knee" remains open. We have, however, also pointed out, how close some of
the most recent observational results are to the corresponding theoretical
flux estimates. This gives the next generation of \gr detectors a decisive
role. Let us see what the experiment says!

\end{document}